\documentclass[a4paper,11pt]{article}
\usepackage{pos}
\usepackage[utf8]{inputenc}
\usepackage{amsmath}
\usepackage{xcolor}
\usepackage{graphicx}
\usepackage{amsfonts}
\usepackage{caption}
\usepackage{subcaption}
\usepackage{tabularx}
\usepackage{multirow}
\usepackage{booktabs}
\usepackage{cleveref}

\title{Position-Space Renormalisation of the Energy-Momentum Tensor}
\ShortTitle{Position-Space Renormalisation of the Energy-Momentum Tensor}

\author*[a]{Henrique Bergallo Rocha}
\author[a]{Luigi Del Debbio}
\author[b,c,d]{Andreas J\"uttner}
\author[b,c,e]{Ben Kitching-Morley}
\author[a]{Joseph K. L. Lee}
\author[a]{Antonin Portelli}
\author[c,e]{Kostas Skenderis}

\affiliation[a]{Higgs Centre for Theoretical Physics, School of Physics and Astronomy, The University of Edinburgh,
Edinburgh EH9 3FD, United Kingdom}
\affiliation[b]{School of Physics and Astronomy, University of Southampton, Southampton SO17 1BJ, United Kingdom}
\affiliation[c]{STAG Research Center, University of Southampton, Highfield, Southampton SO17 1BJ, United Kingdom}
\affiliation[d]{Theoretical Physics Department, CERN, Geneva, Switzerland}
\affiliation[e]{Mathematical Sciences, University of Southampton, Highfield, Southampton SO17 1BJ, United Kingdom}

\emailAdd{h.b.rocha@ed.ac.uk}

\abstract{There is increasing interest in the study of nonperturbative aspects of three-dimensional quantum field theories (QFT). They appear as holographic dual to theories of (strongly coupled) gravity. For instance, in Holographic Cosmology, the two-point function of the Energy-Momentum Tensor (EMT) of a particular class of three-dimensional  QFTs can be mapped into the power spectrum of the Cosmic Microwave Background in the gravitational theory. However, the presence of divergent contact terms poses challenges in extracting a renormalised EMT two-point function on the lattice. Using a $\phi^4$ theory of adjoint scalars valued in the $\mathfrak{su}(N)$ Lie Algebra as a proof-of-concept motivated by Holographic Cosmology, we apply a novel method for filtering out such contact terms by making use of infinitely differentiable "bump" functions which enforce a smooth window that excludes contributions at zero spatial separation. The process effectively removes the local contact terms and allows us to extract the continuum limit behaviour of the renormalised EMT two-point function.}

\FullConference{%
The 39th International Symposium on Lattice Field Theory,\\
8th-13th August, 2022,\\
Rheinische Friedrich-Wilhelms-Universität Bonn, Bonn, Germany
}

\begin{document}
\maketitle

\section{Introduction}

In holographic models of cosmology, the scalar power spectrum of the Cosmic Microwave Background (CMB) is computed from the two-point function of the Energy-Momentum Tensor (EMT) \cite{McFadden2010c}. We can decompose the EMT two-point function as \cite{Coriano2021a} in this case as

\begin{equation}\label{eq:DS}
\langle T_{ij}(\bar q)T_{kl}(-\bar q)\rangle=A(\bar q)\Pi_{ijkl}+B(\bar q) \pi_{ij}\pi_{kl},
\end{equation}
where
\begin{equation}
    \pi_{\mu\nu}=\delta_{\mu\nu}-\frac{q_\mu q_\nu}{q^2}
\end{equation}
is the transverse projector and
\begin{equation}
    \Pi_{\mu\nu\rho\sigma}=\frac{1}{2}(\pi_{\mu \rho} \pi_{\nu \sigma} + \pi_{\mu \sigma} \pi_{\nu \rho} - \pi_{\mu\nu} \pi_{\rho\sigma})
\end{equation}
is the transverse-traceless projector. The $B(q)$ form factor in \cref{eq:DS} maps into the CMB scalar power spectrum via the relation \cite{McFadden2010b,Easther:2011wh}:
 
\begin{equation}
\Delta^2_R(q)=\frac{-q^3}{16 \pi^2 \text{Im} B(-iq)}=\frac{\Delta_0^2}{1+\frac{gq*}{q}\log|\frac{q}{\beta g q*}|}
\end{equation}

Perturbatively, it has been shown in \cite{Afshordi2017a,PhysRevLett.118.041301} that for high multipole momenta ($l\gtrsim 30$) the fit that the model gives to cosmological data is competitive with that of $\Lambda$CDM. The low multipole momentum region, however, maps into the non-perturbative regime of the QFT, and therefore a lattice treatment of it is rendered necessary.

\section{Algorithm}

The lattice simulations discussed in the following sections were written using the Grid library \cite{Boyle2015}, whereupon a Heatbath-Overrelaxation algorithm was used to update the scalar fields. The algorithm consists of a heatbath update followed by $n$ overrelaxation, or reflection, updates. These also contain Metropolis accept/reject steps where appopriate to account for non-gaussianities. These algorithms are discussed in more detail in \cite{Adler1988b}, and the update prescription we used follows closely the one laid out in \cite{Bunk1995a}, with slight alterations for numerical stability and without the gauge updates.

\section{The Discretised Energy-Momentum Tensor and Ward Identities}

Motivated by Holographic Cosmology, we focus on the theory defined by the following lattice action:

\begin{equation}\label{actionscalar}
S=\frac{a^3N}{g}\sum_{x\in\Lambda^3}\textrm{Tr}\left\{\sum_\mu[\Delta_\mu\phi(x)]^2+(m^2-m_c^2) \phi^2(x)+\phi^4(x)\right\}.
\end{equation}
Here, our scalar fields $\phi(x)$ are traceless hermitian $N\times N$ matrices valued in the $\mathfrak{su}(N)$ algebra and $\Delta_\mu$ is the forward discrete derivative. Furthermore, the theory here is presented with large-$N$ scaling. This theory is superrenormalisable and contains a continuum second-order phase transition between a symmetric and a broken phase at $m^2=0$. Furthermore, this theory is perturbatively IR-divergent, but it has been shown by the LatCos collaboration that it is nonperturbatively finite \cite{Cossu2021a}, in addition to the expectation that it should be IR-finite due to its superrenormalisability \cite{Jackiw1981a,PhysRevD.23.2305}. A tentative form of the bare lattice Energy-Momentum Tensor $T^0_{\mu\nu}$ may be obtained by replacing the derivatives of the continuum $T_{\mu\nu}$ with discrete central derivatives, here denoted by $\bar\Delta_\mu$: 
\begin{multline}\label{eq:latticeT}
    T^0_{\mu\nu}=\frac{N}{g}\textrm{Tr}\Bigg\{2(\bar\Delta_\mu\phi)(\bar \Delta_\nu\phi)-\delta_{\mu\nu}\left[\sum_\rho(\bar\Delta_\rho\phi)^2+(m^2-m_c^2)\phi^2+\phi^4\right]\\
    +\xi\left[\delta_{\mu\nu}\sum_\rho(\bar\Delta_\rho \phi)^2-(\bar\Delta_\mu\phi)(\bar\Delta_\nu\phi)\right]\Bigg\}.
\end{multline}

The last term in brackets, which multiplies the constant $\xi$, is the improvement term which accounts for non-minimal coupling of the QFT to gravity. $\xi$ is a parameter to be fixed by comparing with CMB data. The bare lattice $T^0_{\mu\nu}$ as defined in \cref{eq:latticeT}, however, does not satisfy the continuum Ward Identity (WI) due to the breaking of continuum translational symmetry. Explicitly,

\begin{equation} \label{eq:lattice-WI}
  	\langle \bar \Delta_\mu T_{\mu\nu}^0 (x) P(y)\rangle = -\bigg \langle \frac{\delta P(y)}{\delta \phi(x)}\bar \Delta_\nu \phi(x) \bigg\rangle + \langle X_\nu (x) P(y) \rangle,
 \end{equation}
where $P(x)$ and $X_\mu(x)$ are lattice operators. In order to restore the WI as the lattice regulator is removed (i.e. taking $a\rightarrow0$), we require the second term in \cref{eq:lattice-WI} to vanish in this limit. However, due to radiative corrections inducing mixings with lower-dimensional operators than $T^0_{\mu\nu}$, the second term in fact diverges with $a^{-1}$. Therefore, it requires renormalisation. As a renormalisation prescription, we will require that the WI be restored when we remove the regulator by subtracting the divergent contribution from the bare lattice Energy-Momentum Tensor.

It has been shown \cite{Caracciolo1988} that in the 4D theory $T^0_{\mu\nu}$ mixes with 5 different lower-dimensional operators. Power-counting tells us that in the 3D theory there is only one operator with which it mixes. Namely, $\tilde O=\delta_{\mu\nu}\textrm{Tr}\phi^2$. Therefore, we subtract from the bare EMT a divergent term that is proportional to this operator:

\begin{equation}\label{eq:Trenorm}
    T_{\mu\nu}^R=T_{\mu\nu}^0-\frac{Nc_3}{a}\delta_{\mu\nu}\textrm{Tr} \phi^2,
\end{equation}
where $c_3$ is a constant to be determined. Perturbatively, we may obtain it, for instance to one loop:

\begin{equation}
    c_3^{\textrm{1-loop}}=\left(2-\frac{3}{N^2}\right)\left(\frac{6Z_0-1}{12}\right).
\end{equation}
Nevertheless, since we wish to consider the theory at its critical point where the correlation length diverges, we cannot trust perturbative results to be accurate. Therefore, we need to turn to non-perturbative methods. Consider the following insertion of $T^0_{\mu\nu}$:

\begin{equation}\label{eq:CmunuExpr}
    C^0_{\mu\nu} (q) = \frac{N}{g} a^3 \sum_x e^{-i q \cdot x} {\langle T_{\mu\nu}^0 (x) \textrm{Tr} \phi^2 (0)\rangle}  = {C}_{\mu\nu} (q) + \frac{g}{a}c_3\delta_{\mu\nu} C_2 (q) + \frac{\kappa}{a} \delta_{\mu\nu},
\end{equation}
where the $\kappa/a$ factor is a contact term, and
\begin{equation}
    C_2 (q) = \left(\frac{N}{g}\right)^2 a^3 \sum_x e^{-i q \cdot x} {\langle\textrm{Tr}\phi^2(x) \textrm{Tr} \phi^2 (0)\rangle}.
\end{equation}
Furthermore, $C_{\mu\nu}$ is the corresponding finite, continuum correlator and the expression in the last equality \cref{eq:CmunuExpr} has been obtained by inserting \cref{eq:Trenorm} into the definition of $C^0_{\mu\nu}$. Both the contact term and the WI-breaking term diverge with $a^{-1}$, so if we wish to nonperturbatively calculate the value of $c_3$, we need to untangle these two contributions. This can be done by filtering out the contact term in position-space with the use of a window function.

\section{The Window Function}

In order to remove the contributions from contact terms, we introduce the \textit{position-space window function} $\Gamma_{r_0,\epsilon}(x)$, defined as follows:

\begin{equation}
    \Gamma_{r_0,\epsilon}(x)=
    \begin{cases}
    0, & 0<x<r_0\\
    \bar \Gamma_{r_0,\epsilon}(x), & r_0\leq x\leq r_0+\epsilon\\
    1, & r_0+\epsilon\leq x<\infty 
    \end{cases}
\end{equation}
where, between $r_0$ and $\epsilon$, the function is defined as
\begin{equation}
    \bar \Gamma_{r_0,\epsilon}(x)= 1- \frac{\int_x^{r_0+\epsilon}  \beta(u,r_0, \epsilon) \, du}{\int_{r_0}^{r_0+\epsilon} \beta(u,r_0, \epsilon) \, du}
\end{equation}
where here we have
\begin{align}
\beta(x,r_0, \epsilon) &= f(x-r_0) f(r_0 + \epsilon - x),\\
    f(x) &= 
    \begin{cases}
      0 & \text{for } x \leq 0 \\
      \exp{\left(-\frac{1}{x}\right)} & \text{for }  x > 0
    \end{cases}       .
\end{align}
The relevant properties of this construction of $\Gamma_{r_0,\epsilon}(x)$ are as follows:
\begin{enumerate}
    \item It is zero for any value of $x$ less than a minimum radius $r_0$, and therefore will completely exclude any contribution coming from this window.
    \item It is one for any value of $x$ greater than $r_0+\epsilon$, and therefore will not affect contributions to the function beyond this radius.
    \item It smoothly interpolates between 0 and 1 in the window $r_0\leq x \leq r_0+\epsilon$, that is, it is $C^{\infty}$ in this window, and therefore does not generate discontinuity artifacts.
    \item Due to the Paley-Wiener theorem \cite{Wiener1934}, the Fourier Transform  $\mathcal{F}[{\Gamma}_{r_0,\epsilon}(p)]$ decays faster than any power of $\frac{1}{|p|}$, and goes asymptotically as $|p|^{-n} e^{-|p|^m}$ for some $n, m$ for large $|p|$. The parameter $\epsilon$ determines the rate of decay of $\mathcal{F}[{\Gamma}_{r_0,\epsilon}(p)]$, i.e. the larger $\epsilon$ is, or equivalently, the smoother the window function, the more rapid its momentum-space representation decays.
\end{enumerate}

\begin{figure}
    \includegraphics[width=\linewidth]{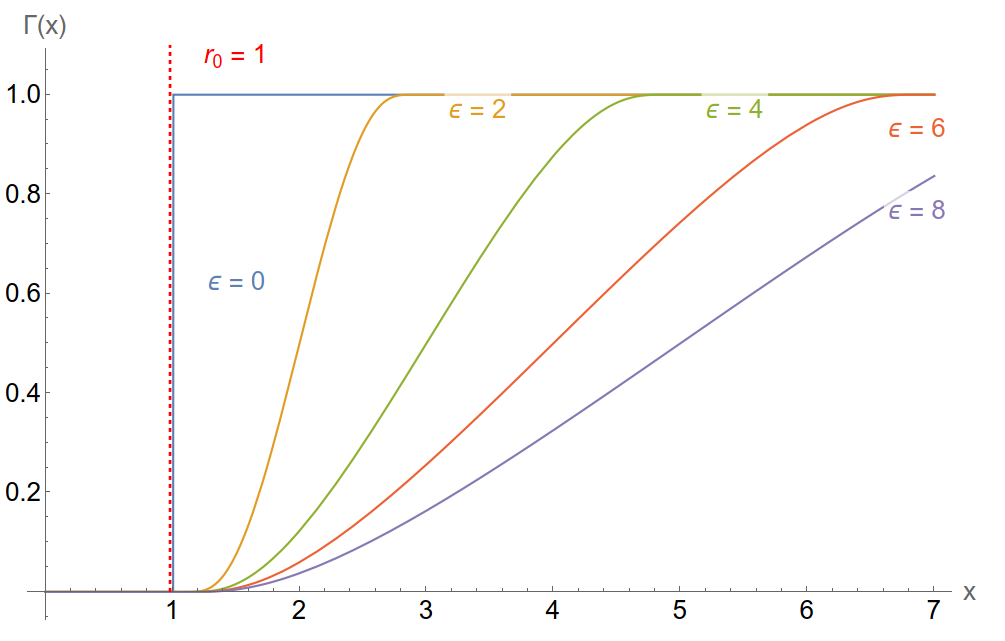}
    \caption{Plot of the position-space window function for different choices of $\epsilon$, with a fixed $r_0=1$.}
\label{fig:windowpos}
\end{figure}

\begin{figure}
    \includegraphics[width=\linewidth]{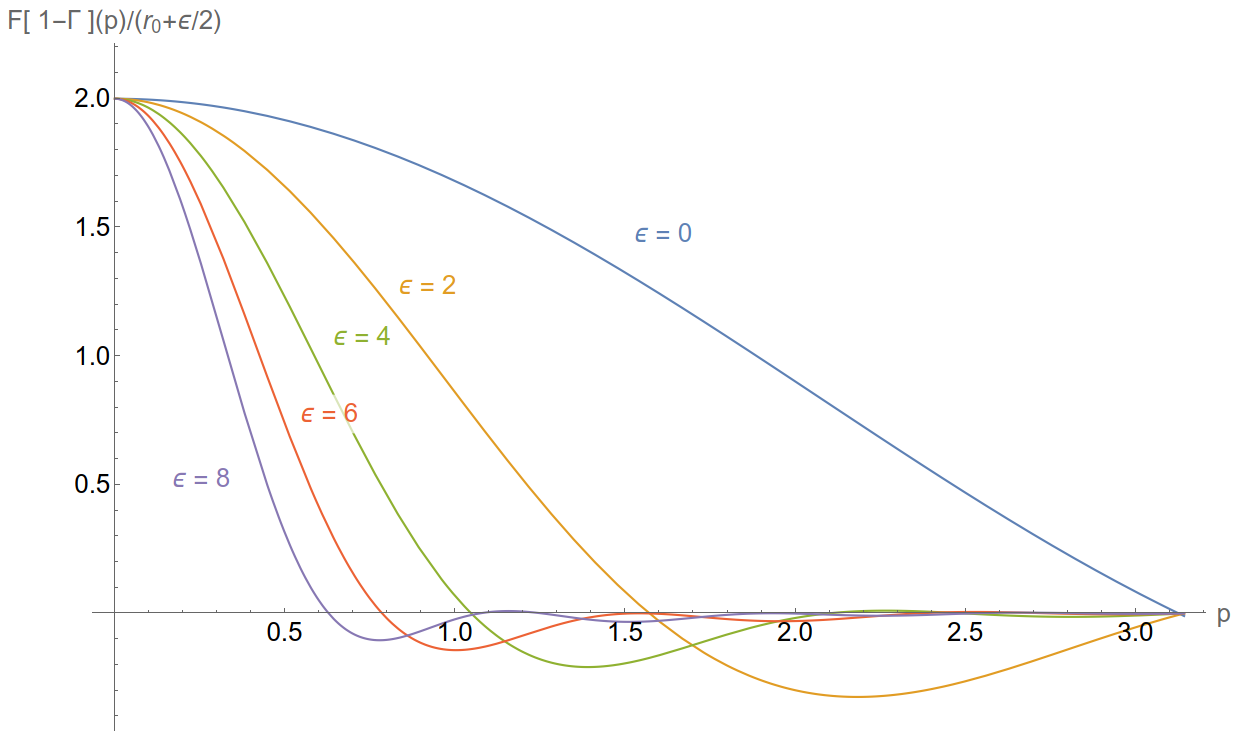}
    \caption{Plot of the Fourier Transform of $1-\Gamma_{r_0,\epsilon}(x)$ for different choices of $\epsilon$, with a fixed $r_0=1$. The subtraction is to restrict the Fourier Transform to a compact domain, and furthermore the curves are divided by a factor of $r_0+\epsilon/2$ for better visualisation.}
\label{fig:windowmom}
\end{figure}

In \cref{fig:windowpos} and \cref{fig:windowmom} it is possible to see the behaviour of the window and its Fourier Transform for different choices of $\epsilon$.

\section{EMT Position-Space Renormalisation}
In order to remove the contact term contributions from a given lattice operator $\mathcal{O}(q)$, we define \textit{windowing} as the following operation:

\begin{equation}
    W_{r_0,\epsilon}[\mathcal{O}](q)=\left(\frac{a}{L}\right)^3\sum_{x}e^{-iq\cdot x}\,\Gamma_{r_0,\epsilon}(|x|)\sum_{q'}e^{iq'\cdot x}\,\mathcal{O}(q').
\end{equation}
This windowing operation is linear, and therefore applying it to \cref{eq:CmunuExpr} gives

\begin{equation}\label{eq:WinCmunuExpr}
    W_{r_0,\epsilon}[C^0_{\mu\nu}] (q) = W_{r_0,\epsilon}[{C}_{\mu\nu}](q) + \frac{g}{a}c_3\delta_{\mu\nu}W_{r_0,\epsilon}[C_2] (q) + W_{r_0,\epsilon}\left[\frac{\kappa}{a} \delta_{\mu\nu}\right].
\end{equation}
The last term in \cref{eq:WinCmunuExpr} is a contact term and therefore yields zero when windowed. Rearranging this expression to isolate $c_3$,

\begin{equation}
c_3=\frac{a}{g}\left( \frac{W_{r_0,\epsilon}[C^0_{\mu\nu}](q)-W_{r_0,\epsilon}[{C}_{\mu\nu}](q)}{W_{r_0,\epsilon}[C_2](q)} \right).
\end{equation}
Restricting ourselves to the zero mode ($q=0$) and taking the limit $\epsilon\rightarrow\infty$, it is possible to show that this expression behaves as

\begin{equation}
c_3\sim\frac{a}{g}\left( \frac{C^0_{\mu\nu}(0)}{C_2(0)}-\frac{b_2}{\epsilon} \right)
\end{equation}
where $b_2$ is some constant. This suggests that we can vary $\epsilon$ while measuring the ratio between the bare correlators and fit these results to the form 
\begin{equation}\label{eq:c3Ratio}
  \frac{a W_{r_0, \epsilon}[C^0_{22}] (q_l=0)}{W_{r_0, \epsilon}[C_2] (q_l=0)} = \bar{c}_3 + \frac{b}{\epsilon},
\end{equation}
where $b$ and $\bar {c}_3$ are fit parameters, whence we can extrapolate to find $c_3$ in the $1/\epsilon\rightarrow0$ limit, as the left-hand side of \cref{eq:c3Ratio} contains only lattice observables.

\section{EMT Renormalisation Results}

\begin{figure}
  \begin{subfigure}[t]{.5\textwidth}
	% include first image
	\includegraphics[width=1.0\linewidth]{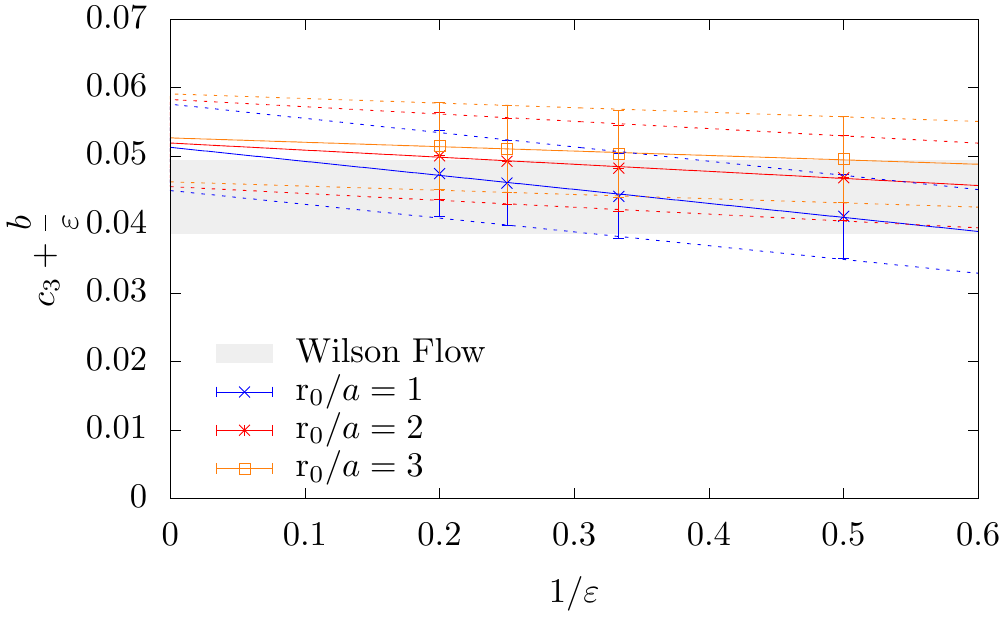}  
	\caption{$ag=0.1, (am)^2=-0.0313$}
  \end{subfigure}
  \begin{subfigure}[t]{.5\textwidth}
	% include second image
	\includegraphics[width=1.0\linewidth]{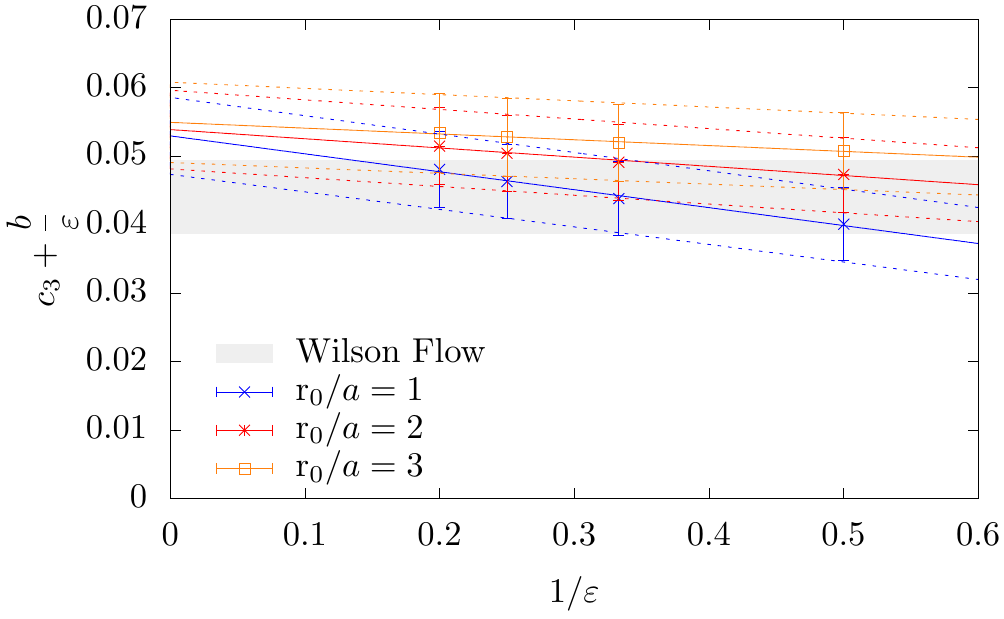}
	\caption{$ag=0.2, (am)^2=-0.06215$}
  \end{subfigure}\\
  \par\bigskip\centering
  \begin{subfigure}[t]{.5\textwidth}
    % include second image
    \includegraphics[width=1.0\linewidth]{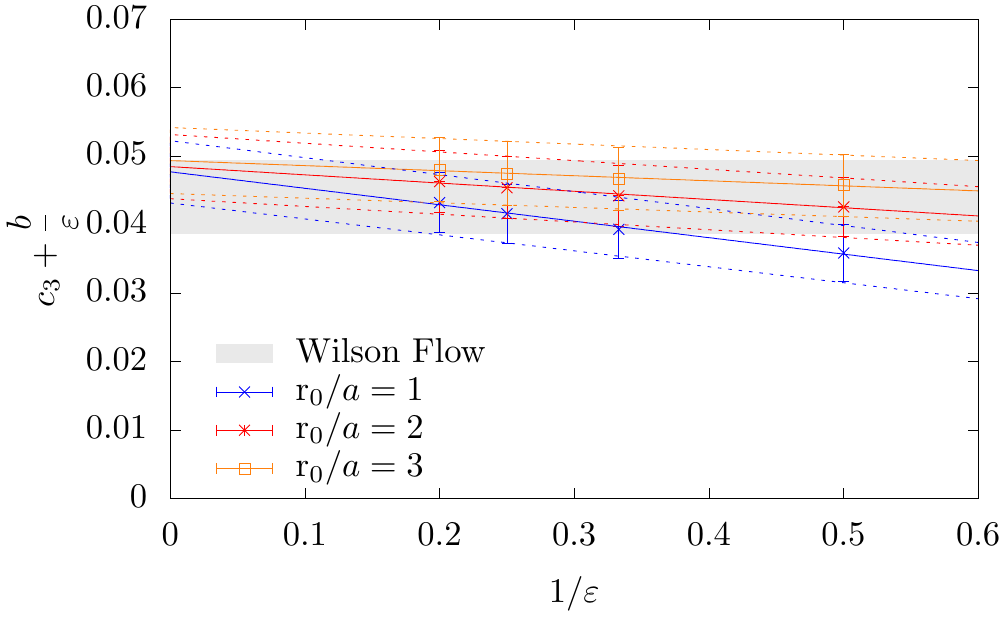}
    \caption{$ag=0.3, (am)^2=-0.09275$}
  \end{subfigure}
  
  \caption{Fit results for the renormalisation constant $\bar{c}_3$ using the ansatz \cref{eq:c3Ratio}, which is the value of the $y$ intercept. This is performed for three ensembles ($N_L=256$) and different choices of window radius $r_0/a \in \left\{1,2,3\right\}$, as represented by the three colours. The grey bands correspond to the result for $c_3$ obtained using the Wilson flow.}
  \label{fig:c3_window_fit_epsilon}
\end{figure}

On \cref{fig:c3_window_fit_epsilon} it is possible to see the results of the fit given by the ansatz in \cref{eq:c3Ratio} and their extrapolation to $\epsilon\rightarrow\infty$ for three different choices of $ag$. The simulated masses were in the vicinity of the critical point. It can be seen that the extrapolated value of $c_3$, given by the $y$-intercept, gives overlap between the error bands between the position-space method and the Wilson Flow method, as obtained in \cite{DelDebbio2020a}.  

\section{Two-Point Function Renormalisation on Synthetic Data}

\begin{figure}
    
  \begin{subfigure}[t]{.5\textwidth}
    \centering
    % include first image
    \colorbox{white}{
    \includegraphics[width=0.9\linewidth]{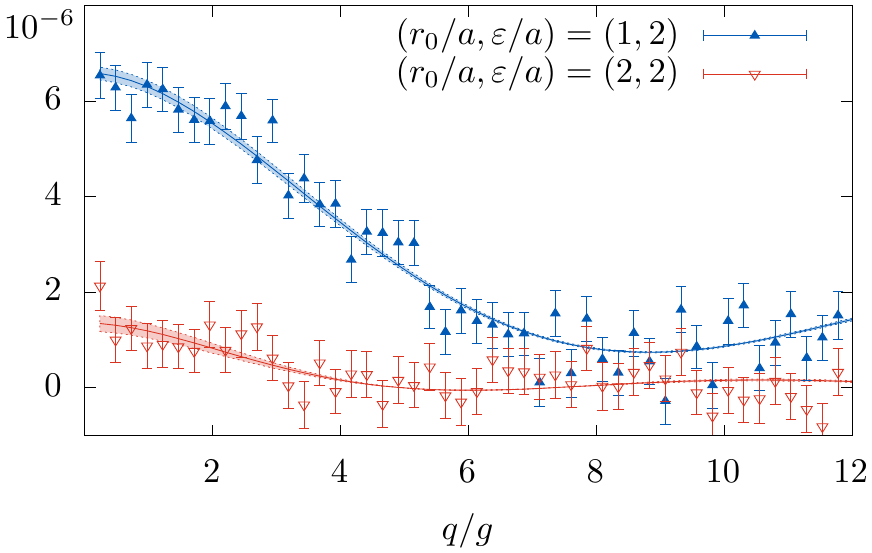}  
    }
    \caption{$\sigma = 1$}
  \end{subfigure}
  \begin{subfigure}[t]{.5\textwidth}
    \centering
    % include first image
    \colorbox{white}{
    \includegraphics[width=0.9\linewidth]{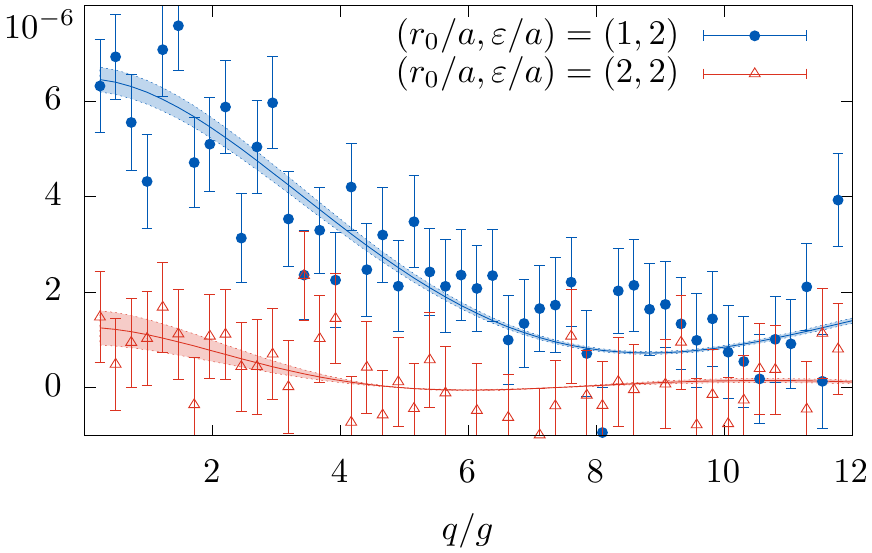}  
    }
    \caption{$\sigma = 2$}
  \end{subfigure}\\
  \par\bigskip
  \begin{subfigure}[t]{.5\textwidth}
    \centering
    % include first image
    \colorbox{white}{
    \includegraphics[width=0.9\linewidth]{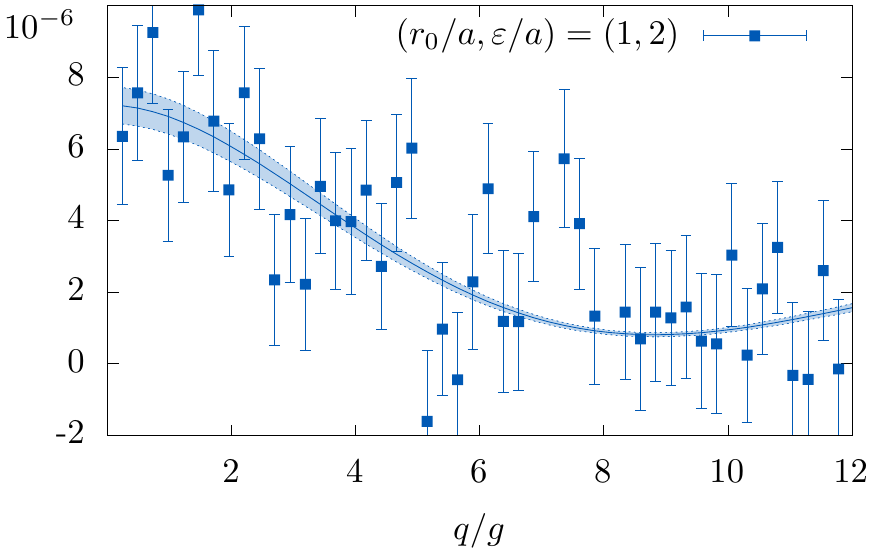}  
    }
    \caption{$\sigma = 4$}
  \end{subfigure}
  \begin{subfigure}[t]{.5\textwidth}
    \centering
    % include first image
    \colorbox{white}{
    \includegraphics[width=0.9\linewidth]{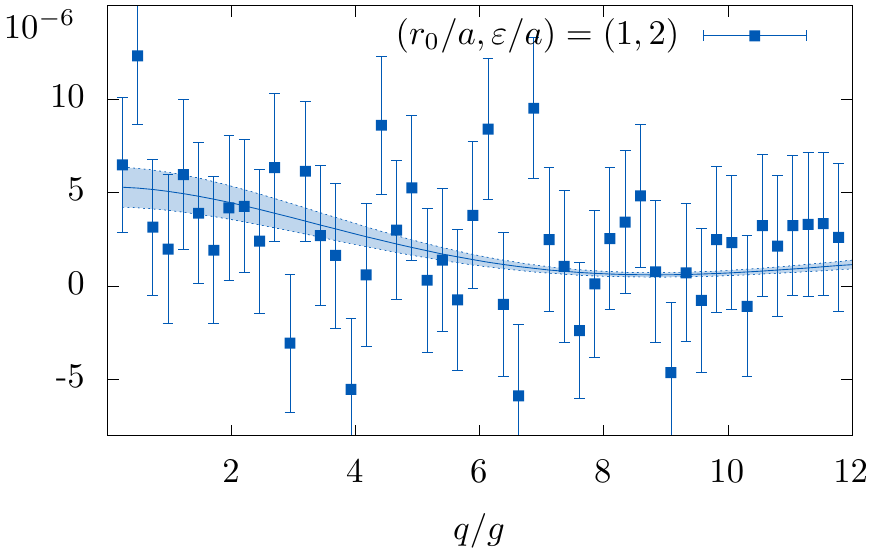}  
    }
    \caption{$\sigma = 8$}
  \end{subfigure}
  \caption{Fits against synthetic data with $\alpha_0=\beta_0=\gamma_0=0.01$ for increasing levels of added noise $\sigma$. The figures with $\sigma=1$ and $\sigma=2$ also show the fits for $(r_0/a,\epsilon/a)=(2,2)$ since this window choice still contains some nonzero signal.}
\label{fig:synthdata}
  \end{figure}

In the continuum, it is expected that the two-point function of the EMT will contain terms proportional to $(q/g)^3$ and $(q/g)^2\log(q/g)$, in addition to contact terms. In principle, the windowing procedure can remove such contact terms from lattice data such that we are left only with the signal whence we may extract the form factors $A(q)$ and $B(q)$ from \cref{eq:DS}. As a proof-of-concept, we generate synthetic data according to the distribution

\begin{equation}
    \frac{C(q)}{g^3}= \alpha_0 \left(\frac{\hat{q}}{g}\right)^3 + \frac{\beta_0}{ag} \left(\frac{\hat{q}}{g}\right)^2 + \frac{\gamma_0}{(ag)^3},
\end{equation}
where $\alpha_0$, $\beta_0$, and $\gamma_0$ are parameters to be chosen. Only the first term on the right-hand side contains the relevant signal, as the other two are contact terms. To this generated momentum distribution, Gaussian noise with standard deviation $\sigma$ is added. The resulting function is then windowed, and subsequently fitted against a windowed "pure" $\hat q^3$ distribution, with the intent of recovering the value of $\alpha_0$. Some of those fits are shown in \cref{fig:synthdata}.

The results of some such fits are given on \cref{table:synthfits}.

\begin{table}
  \centering
          \begin{tabular}{cclll}
            \toprule
              \multirow{2}{*}{$\sigma$}&\multirow{2}{*}{$r_0/a$}&\multicolumn{3}{c}{$\epsilon/a$}\\\cmidrule(lr){3-5}
             &&\multicolumn{1}{c}{1}           & \multicolumn{1}{c}{2}           & \multicolumn{1}{c}{3}         \\ \midrule
             \multirow{4}{*}{1}&1 & 0.0100(1)   & 0.0097(2)   & 0.0103(4)   \\
             &2 & 0.0106(9)   & 0.0117(15)  & 0.0063(24)           \\
             &3 & 0.0095(33)  & 0.0126(49)  & 0.0048(75)   \\
             &4 & -0.0001(87) & 0.0143(121) & 0.0208(158)   \\    \midrule
             \multirow{4}{*}{2}&1 & 0.0103(2)   & 0.0096(4)    & 0.0105(8)   \\
             &2 & 0.0120(20)  & 0.0109(31)   & 0.0104(48)    \\
             &3 & 0.0141(65)  & 0.0020(106)  & 0.0195(143)    \\
             &4 & 0.0010(166) & -0.0057(243) & -0.0154(319)  \\ \midrule
             \multirow{4}{*}{4}&1 & 0.0098(3)    & 0.0107(8)    & 0.0094(16) \\
             &2 & 0.0142(38)   & 0.0072(62)   & 0.0142(97)    \\
             &3 & 0.0095(129)  & -0.0176(196) & 0.0448(286)     \\
             &4 & -0.0133(345) & 0.0078(458)  & -0.1224(647)   \\
              \bottomrule
          \end{tabular}
          
  \caption{Some results of the best fit for $\alpha_0$ for various windows and noise magnitudes. For sufficiently small windows, the fit accurately recovers the original value of $\alpha_0=0.01$.}
  \label{table:synthfits}
\end{table}

\section{Conclusion}
With the position-space method, we have managed to renormalise the EMT and obtain results that are compatible with those yielded by the Wilson Flow method. Furthermore, the method can also in principle get rid of contact term contributions in the two-point function to recover the continuum correlator parameters. We intend to show in future work that such a method can also work on real lattice data and to apply it to other more complex theories, like one containing gauge fields alongside scalars. Thus, this method may pave the way for allowing tests of the predictions of Holographic Cosmology nonperturbatively.

\section{Acknowledgements}
A. J. and K. S. acknowledge funding from STFC consolidated grants ST/ P000711/1 and ST/T000775/1. A.P. is supported in part by UK STFC grant ST/P000630/1. A.P. also received funding from the European Research Council (ERC) under the European Union’s Horizon 2020 research and innovation programme under grant agreements No 757646 \& 813942. J. K. L. L., and H. B. R are funded in part by the European Research Council (ERC) under the European Unions Horizon 2020 research and innovation programme under Grant Agreement No. 757646. J. K. L. L. is also partly funded by the Croucher Foundation through the Croucher Scholarships for Doctoral Study. B. K. M. was supported by the EPSRC Centre for Doctoral Training in Next Generation Computational Modelling Grant No. EP/L015382/1. L. D. D. is supported by an STFC Consolidated Grant, ST/ P0000630/1, and a Royal Society Wolfson Research Merit Award, WM140078. Simulations produced for this work were performed using the Grid Library, which is free software under GPLv2. This work was performed using the Cambridge Service for Data Driven Discovery (CSD3), part of which is operated by the University of Cambridge Research Computing on behalf of the STFC DiRAC HPC Facility. The DiRAC component of CSD3 was funded by BEIS capital funding via STFC capital grants ST/P002307/1 and ST/R002452/1 and STFC operations grant ST/R00689X/1. DiRAC is part of the National e-Infrastructure.

\bibliographystyle{plainnat}
\bibliography{Lattice2022.bib}

\begin{thebibliography}{15}
\providecommand{\natexlab}[1]{#1}
\providecommand{\url}[1]{\texttt{#1}}
\expandafter\ifx\csname urlstyle\endcsname\relax
  \providecommand{\doi}[1]{doi: #1}\else
  \providecommand{\doi}{doi: \begingroup \urlstyle{rm}\Url}\fi

\bibitem[Adler(1988)]{Adler1988b}
Stephen~L. Adler.
\newblock {Overrelaxation algorithms for lattice field theories}.
\newblock \emph{Physical Review D}, 37\penalty0 (2):\penalty0 458--471, jan
  1988.
\newblock ISSN 0556-2821.
\newblock \doi{10.1103/PhysRevD.37.458}.
\newblock URL
  \url{https://journals.aps.org/prd/abstract/10.1103/PhysRevD.37.458
  https://link.aps.org/doi/10.1103/PhysRevD.37.458}.

\bibitem[Afshordi et~al.(2017{\natexlab{a}})Afshordi, Corian\`o, Delle~Rose,
  Gould, and Skenderis]{PhysRevLett.118.041301}
Niayesh Afshordi, Claudio Corian\`o, Luigi Delle~Rose, Elizabeth Gould, and
  Kostas Skenderis.
\newblock From planck data to planck era: Observational tests of holographic
  cosmology.
\newblock \emph{Phys. Rev. Lett.}, 118:\penalty0 041301, Jan
  2017{\natexlab{a}}.
\newblock \doi{10.1103/PhysRevLett.118.041301}.
\newblock URL \url{https://link.aps.org/doi/10.1103/PhysRevLett.118.041301}.

\bibitem[Afshordi et~al.(2017{\natexlab{b}})Afshordi, Gould, and
  Skenderis]{Afshordi2017a}
Niayesh Afshordi, Elizabeth Gould, and Kostas Skenderis.
\newblock {Constraining holographic cosmology using Planck data}.
\newblock \emph{Physical Review D}, 95\penalty0 (12):\penalty0 1--25,
  2017{\natexlab{b}}.
\newblock ISSN 24700029.
\newblock \doi{10.1103/PhysRevD.95.123505}.

\bibitem[Appelquist and Pisarski(1981)]{PhysRevD.23.2305}
Thomas Appelquist and Robert~D. Pisarski.
\newblock High-temperature yang-mills theories and three-dimensional quantum
  chromodynamics.
\newblock \emph{Phys. Rev. D}, 23:\penalty0 2305--2317, May 1981.
\newblock \doi{10.1103/PhysRevD.23.2305}.
\newblock URL \url{https://link.aps.org/doi/10.1103/PhysRevD.23.2305}.

\bibitem[Boyle et~al.(2016)Boyle, Cossu, Yamaguchi, and Portelli]{Boyle2015}
Peter~A. Boyle, Guido Cossu, Azusa Yamaguchi, and Antonin Portelli.
\newblock {Grid: A next generation data parallel C++ QCD library}.
\newblock \emph{Proceedings of The 33rd International Symposium on Lattice
  Field Theory — PoS(LATTICE 2015)}, \penalty0 (July):\penalty0 023, jul
  2016.
\newblock \doi{10.22323/1.251.0023}.
\newblock URL \url{https://pos.sissa.it/251/023}.

\bibitem[Bunk(1995)]{Bunk1995a}
B.~Bunk.
\newblock {Monte-Carlo methods and results for the electro-weak phase
  transition}.
\newblock \emph{Nuclear Physics B (Proceedings Supplements)}, 42\penalty0
  (1-3):\penalty0 566--568, 1995.
\newblock ISSN 09205632.
\newblock \doi{10.1016/0920-5632(95)00313-X}.

\bibitem[Caracciolo et~al.(1988)Caracciolo, Curci, Menotti, and
  Pelissetto]{Caracciolo1988}
Sergio Caracciolo, Giuseppe Curci, Pietro Menotti, and Andrea Pelissetto.
\newblock {The energy-momentum tensor on the lattice: The scalar case}.
\newblock \emph{Nuclear Physics, Section B}, 309\penalty0 (4):\penalty0
  612--624, 1988.
\newblock ISSN 05503213.
\newblock \doi{10.1016/0550-3213(88)90332-X}.

\bibitem[Corian{\`{o}} et~al.(2021)Corian{\`{o}}, {Delle Rose}, and
  Skenderis]{Coriano2021a}
Claudio Corian{\`{o}}, Luigi {Delle Rose}, and Kostas Skenderis.
\newblock {Two-point function of the energy-momentum tensor and generalised
  conformal structure}.
\newblock \emph{European Physical Journal C}, 81\penalty0 (2):\penalty0 1--33,
  2021.
\newblock ISSN 14346052.
\newblock \doi{10.1140/epjc/s10052-021-08892-5}.
\newblock URL \url{https://doi.org/10.1140/epjc/s10052-021-08892-5}.

\bibitem[Cossu et~al.(2021)Cossu, {Del Debbio}, J{\"{u}}ttner, Kitching-Morley,
  Lee, Portelli, Rocha, and Skenderis]{Cossu2021a}
Guido Cossu, Luigi {Del Debbio}, Andreas J{\"{u}}ttner, Ben Kitching-Morley,
  Joseph~K.L. Lee, Antonin Portelli, Henrique~Bergallo Rocha, and Kostas
  Skenderis.
\newblock {Nonperturbative Infrared Finiteness in a Superrenormalizable Scalar
  Quantum Field Theory}.
\newblock \emph{Physical Review Letters}, 126\penalty0 (22):\penalty0 221601,
  jun 2021.
\newblock ISSN 0031-9007.
\newblock \doi{10.1103/PhysRevLett.126.221601}.
\newblock URL \url{https://link.aps.org/doi/10.1103/PhysRevLett.126.221601}.

\bibitem[{Del Debbio} et~al.(2021){Del Debbio}, Dobson, J{\"{u}}ttner,
  Kitching-Morley, Lee, Nourry, Portelli, {Bergallo Rocha}, and
  Skenderis]{DelDebbio2020a}
Luigi {Del Debbio}, Elizabeth Dobson, Andreas J{\"{u}}ttner, Ben
  Kitching-Morley, Joseph~K.L. Lee, Valentin Nourry, Antonin Portelli, Henrique
  {Bergallo Rocha}, and Kostas Skenderis.
\newblock {Renormalization of the energy-momentum tensor in three-dimensional
  scalar SU(N) theories using the Wilson flow}.
\newblock \emph{Physical Review D}, 103\penalty0 (11):\penalty0 114501, jun
  2021.
\newblock ISSN 2470-0010.
\newblock \doi{10.1103/PhysRevD.103.114501}.
\newblock URL \url{https://link.aps.org/doi/10.1103/PhysRevD.103.114501}.

\bibitem[Easther et~al.(2011)Easther, Flauger, McFadden, and
  Skenderis]{Easther:2011wh}
Richard Easther, Raphael Flauger, Paul McFadden, and Kostas Skenderis.
\newblock {Constraining holographic inflation with WMAP}.
\newblock \emph{JCAP}, 09:\penalty0 030, 2011.
\newblock \doi{10.1088/1475-7516/2011/09/030}.

\bibitem[Jackiw and Templeton(1981)]{Jackiw1981a}
R.~Jackiw and S.~Templeton.
\newblock {How super-renormalizable interactions cure their infrared
  divergences}.
\newblock \emph{Physical Review D}, 23\penalty0 (10):\penalty0 2291--2304,
  1981.
\newblock ISSN 05562821.
\newblock \doi{10.1103/PhysRevD.23.2291}.

\bibitem[McFadden and Skenderis(2010{\natexlab{a}})]{McFadden2010b}
Paul McFadden and Kostas Skenderis.
\newblock {The Holographic Universe}.
\newblock \penalty0 (i), jan 2010{\natexlab{a}}.
\newblock \doi{10.1088/1742-6596/222/1/012007}.
\newblock URL \url{http://arxiv.org/abs/1001.2007
  http://dx.doi.org/10.1088/1742-6596/222/1/012007}.

\bibitem[McFadden and Skenderis(2010{\natexlab{b}})]{McFadden2010c}
Paul McFadden and Kostas Skenderis.
\newblock {Holography for cosmology}.
\newblock \emph{Physical Review D - Particles, Fields, Gravitation and
  Cosmology}, 81\penalty0 (2), 2010{\natexlab{b}}.
\newblock ISSN 15507998.
\newblock \doi{10.1103/PhysRevD.81.021301}.

\bibitem[Wiener and Paley(1934)]{Wiener1934}
N.~Wiener and R.C. Paley.
\newblock \emph{{Fourier transforms in the complex domain}}.
\newblock American Mathematical Society, 1934.
\newblock ISBN 978-0-8218-1019-4.
\newblock URL \url{https://bookstore.ams.org/coll-19}.

\end{thebibliography}

\end{document}